\documentclass[aps,prd,preprintnumbers,nofootinbib,superscriptaddress,amsmath]{revtex4-1}

\usepackage{hyperref}
\usepackage{graphicx}
\usepackage{dcolumn}
\usepackage{bm}
\usepackage[normalem]{ulem}
\usepackage{booktabs}

\usepackage[table]{xcolor}
\definecolor{lightgray}{gray}{0.9} 
\def\mystrut{\vrule height 9.3pt depth 3.1pt width 0pt}

\def\Msun{\mathrm{M}_\odot}

\begin{document}

\title{Primordial Black Hole clusters, phenomenology \& implications}

\author{José Francisco Nuño Siles}\email[]{jose.nunno@uam.es}
\author{Juan García-Bellido}\email[]{juan.garciabellido@uam.es}

\affiliation{Instituto de F\'isica Te\'orica UAM-CSIC, Universidad Aut\'onoma de Madrid, Cantoblanco 28049 Madrid, Spain}

\date{\today}

\preprint{IFT-UAM/CSIC-24-61}

\begin{abstract}
We present direct N-body simulations of black-hole-only clusters with up to $2 \cdot 10^4$ compact objects, zero natal spin and no primordial binaries as predicted by various primordial black hole (PBH) Dark Matter models. The clusters' evolution is computed using {\tt NBODY6++GPU}, including the effects of the tidal field of the galaxy, the kicks of black hole mergers and orbit-averaged energy loss by gravitational radiation of binaries. We investigate clusters with four initial mass distributions, three of which attempt to model a generic PBH scenario using a lognormal mass distribution and a fourth one that can be directly linked to a monochromatic PBH scenario when accretion is considered. More specifically, we dive into the clusters' internal dynamics, describing their expansion and evaporation, along with the resultant binary black hole mergers. We also compare several simulations with and without black hole merger kicks and find modelling implications for the probability of hierarchical mergers.

\end{abstract}

\maketitle


\section{Introduction}

Black holes (BHs) have long captured the attention of astrophysicists due to their elusive and mysterious nature. Recent progress in observational astronomy has revolutionized our ability to study them across various scales, from detailed images of supermassive BHs at the centres of galaxies~\cite{Akiyama_2019} to the detection of microlensing events caused by BHs passing in front of background sources~\cite{GAIA}. More relevant to this work, the detection of gravitational waves emitted by merging binary black holes (BBHs) in Earth-based interferometers~\cite{GWTC3} has improved our understanding of their formation mechanisms and population statistics while, at the same time, giving rise to new fundamental questions~\cite{GW190425,GW190814,GW190521,Clesse:2020ghq}, like the nature of Dark Matter~\cite{Bird:2016dcv,Clesse:2016vqa,Garcia-Bellido:2017fdg}, and the thermal evolution of the early Universe~\cite{Carr:2019kxo}.

In this context, the investigation of black hole-only clusters takes on particular significance as a possible alternative origin capable of shedding light onto some observations. Theoretical considerations, supported by numerical simulations, have demonstrated the plausibility of forming such clusters. For instance, BHs created during the radiation-dominated era are naturally clustered~\cite{Chisholm_2006,Chisholm_2011} if they derive from large non-Gaussian tails~\cite{Ezquiaga:2019ftu,Ezquiaga:2022qpw,Animali:2024jiz}. Other phenomena such as the appearance of closed domain walls~\cite{Belotsky_2019} and their collapse~\cite{PBH_Clusters_1} can lead to their clustering too. We should also mention the possibility of the existence of black-hole-only clusters originating via stellar evolution~\cite{Gieles_2021} for which our analysis and conclusions remain valid with the appropriate time delay.

Despite the progress in the field, our comprehension of the dynamics of black hole-only clusters remains limited~\cite{Trashorras:2020mwn}. The gravitational interactions between thousands of individual BHs together with the influence of the surrounding environment, present hard challenges for any attempt at theoretical modelling given the vast range of scales needed to be considered. However, these complexities also offer a unique opportunity for N-body simulations to excel.

N-body simulations have proven instrumental in understanding a diverse range of complex astrophysical phenomena. Using numerical methods to integrate the gravitational equations of motion for all of the individual particles, we obtain a detailed description of the dynamics governing the evolution of such complex systems~\cite{Inman:2019wvr}. A complete introduction to the theoretical foundations of the field of N-body simulations can be found in Ref.~\cite{aarseth_2003} and a shorter, but less recent one, in~\cite{SPURZEM1999407}.

In this paper, we present direct N-body simulations of black hole-only dense clusters, focusing on systems with up to $2 \cdot10^4$ compact objects. Our simulations incorporate physical effects such as the tidal field of a given host galaxy, black hole relativistic merger kicks, and orbit-averaged energy loss due to gravitational radiation from binary systems. By utilizing the {\tt NBODY6++GPU} code~\cite{2015MNRAS.450.4070W}, we accurately model the evolution of all the BHs for a Hubble time.
The text is structured as follows, in the first section~\ref{sec:Method} we describe the initial conditions in detail, in the second section we analyze the cluster dynamics, trying to understand why the clusters dissolve and expand. The following section is focused on binary black hole mergers and the parameters characterizing these events. In the last section, we summarize our findings and conclude the work.

The main body of this work consists of the description of the physical variables of interest regarding such hypothetical celestial bodies and their behaviour in a Hubble time. We find promising hints of results that could be linked with current observations but would need more in-depth research before satisfying conclusions could be drawn. While there are possibly many more findings in the data, we don't embark on such extensive endeavour and treat this work as a comprehensive proof of concept for future works to be carried out. 

\section{Methodology and initial mass function}\label{sec:Method}

\begin{figure}
    \centering
    \includegraphics[width=0.99\textwidth]{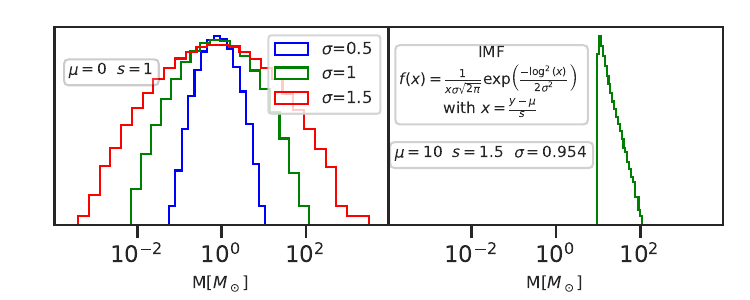}
    \caption{Initial mass function of the clusters. Those based on a typical PBH mass function ($\sigma_i$) are in the left panel, while the monochromatic with accretion IMF ($M\&A$) is in the right one. We also write the general formula for all the IMFs and the values for the specific parameters to reproduce our results. }
    \label{fig:LogNormalDist}
\end{figure}

To evolve our BHs-only clusters, we made use of the direct numerical integrator {\tt NBODY6++GPU}. The models we evolve consist of N black holes, where N ranges from $\mathcal{O}(10^3)$ to $\mathcal{O}(2\cdot10^4)$, drawn randomly from 4 different initial mass functions plotted in Fig~\ref{fig:LogNormalDist}. The exact numbers can be found in Table~\ref{table:ClustersTable}. The first one, a log-normal distribution with $\mu=10$, $s=1.5$ and $\sigma=0.954$ is depicted in the right panel and models, for instance, a monochromatic mass function when accretion is considered and thus the masses of the BHs can increase. The other three represent an approximation to the PBH Thermal model~\cite{Carr:2019kxo,LISACosmologyWorkingGroup:2023njw} composed of a wide mass function with three widths for the log-normal $\sigma=\{0.5,1,1.5\}$. See~\cite{Carr:2023tpt} for a recent review of the model. We will refer to them as \{$M\&A$,$\sigma_{0.5}$,$\sigma_{1}$,$\sigma_{1.5}$\} respectively. We set the initial fraction of binaries as well as the individual spins of the BHs to zero according to theoretical expectations. The BHs are then spatially distributed in such a way that they follow a Plummer distribution~\cite{Plummer} with a Plummer Radius $r_p=10~\mathrm{pc}$. Based on the assumption that they could reside in the halo of a galaxy similar to our own, these clusters are themselves immersed in a central gravitational potential with orbital radius $R_c=34~\mathrm{kpc}$ and central mass $M=4.37\times10^{10}~\Msun$ throughout the entire evolution. This is just a point mass approximation which leads to a circular movement of period $T=2.81\mathrm{Gyr}$. In Table~\ref{table:ClustersTable} we explicitly write key statistics for the individual clusters.

 \begin{table*}
    \centering
    \begin{ruledtabular}\begin{tabular}{l c c c c c | l c c c c c c c c c c c | c c c c c c}
ID  & $M_{\mathrm{total}}[M_\odot]$ & $M_{\mathrm{max}}[M_\odot]$ & $M_{\mathrm{max}}[M_\odot]$ & $R_{\mathrm{HM}}[pc]$ & $R_{\mathrm{HM}}[pc]$ & ID & $M_{\mathrm{total}}[M_\odot]$ & $M_{\mathrm{max}}[M_\odot]$ & $M_{\mathrm{max}}[M_\odot]$ & $R_{\mathrm{HM}}[pc]$ & $R_{\mathrm{HM}}[pc]$  \\[0.25mm] 
  M\&A & $t=0$ & $t=0$ & $t=T_U$ &  $t=0$ & $t=T_U$ & $\sigma_{0.5}$ &$t=0$ & $t=0$ & $t=T_U$ &  $t=0$ & $t=T_U$\\ \hline\\[-3mm]
1295 & 16140 & 46.54 & 46.54 & 2.20 & 27.04 & 1520 & 1389 & 5.03 & 5.03 & 9.72 & 12.25\\[0.5mm]
\makebox[0pt][l]{\fboxsep0pt\colorbox{lightgray}{\mystrut\hspace*{1.0\linewidth}}}\!\!
2570 & 32166 & 66.10 & 97.80 & 2.38 & 27.17 & 3345 & 3007 & 4.50 & 5.08 & 3.23 & 15.55\\[0.5mm]
4046 & 50088 & 63.64 & 91.98 & 4.27 & 26.74 & 5480 & 5009 & 6.67 & 6.67 & 0.77 & 16.21\\[0.5mm]
\makebox[0pt][l]{\fboxsep0pt\colorbox{lightgray}{\mystrut\hspace*{1.0\linewidth}}}\!\!
5199 & 64280 & 69.98 & 74.80 & 2.20 & 28.39 & 7678 & 7008 & 5.69 & 7.52 & 1.92 & 15.15\\[0.5mm]
8077 & 100116 & 78.27 & 78.27 & 12.53 & 29.50 & 9937 & 9001 & 5.48 & 5.48 & 1.25 & 15.85\\[0.5mm]
\makebox[0pt][l]{\fboxsep0pt\colorbox{lightgray}{\mystrut\hspace*{1.0\linewidth}}}\!\!
8922 & 110196 & 76.15 & 81.52 & 13.31 & 32.01 & 12201 & 11001 & 5.00 & 7.77 & 2.90 & 15.29\\[0.5mm]
10372 & 128337 & 104.26 & 104.26 & 6.20 & 27.54 & 14366 & 13005 & 5.42 & 8.22 & 2.65 & 16.04\\[0.5mm]
\makebox[0pt][l]{\fboxsep0pt\colorbox{lightgray}{\mystrut\hspace*{1.0\linewidth}}}\!\!
10535 & 130198 & 68.50 & 74.35 & 9.72 & 28.20 & 16428 & 14912 & 7.72 & 7.72 & 0.48 & 16.04\\[0.5mm]
16159 & 200392 & 57.75 & 113.59 & 4.44 & 27.24 & 18776 & 17007 & 5.40 & 6.51 & 2.29 & 15.00\\[0.5mm]
\makebox[0pt][l]{\fboxsep0pt\colorbox{lightgray}{\mystrut\hspace*{1.0\linewidth}}}\!\!
20738 & 256346 & 134.10 & 145.70 & 1.37 & 26.65 & 20866 & 18918 & 5.31 & 7.36 & 3.91 & 15.35\\[0.5mm]
\hline\\[-3mm]
ID  & $M_{\mathrm{total}}[M_\odot]$ & $M_{\mathrm{max}}[M_\odot]$ & $M_{\mathrm{max}}[M_\odot]$ & $R_{\mathrm{HM}}[pc]$ & $R_{\mathrm{HM}}[pc]$ & ID & $M_{\mathrm{total}}[M_\odot]$ & $M_{\mathrm{max}}[M_\odot]$ & $M_{\mathrm{max}}[M_\odot]$ & $R_{\mathrm{HM}}[pc]$ & $R_{\mathrm{HM}}[pc]$  \\[0.25mm] 
  $\sigma_{1}$ & $t=0$ & $t=0$ & $t=T_U$ &  $t=0$ & $t=T_U$ & $\sigma_{1.5}$ &$t=0$ & $t=0$ & $t=T_U$ &  $t=0$ & $t=T_U$\\ \hline\\[-3mm]
1505 & 2003 & 32.07 & 32.07 & 1.93 & 27.07 & 1220 & 3020 & 63.67 & 63.67 & 3.37 & 58.92\\[0.5mm]
\makebox[0pt][l]{\fboxsep0pt\colorbox{lightgray}{\mystrut\hspace*{1.0\linewidth}}}\!\!
3090 & 4004 & 18.56 & 18.56 & 7.52 & 22.21 & 3423 & 8010 & 105.74 & 105.74 & 7.15 & 41.00\\[0.5mm]
5288 & 7007 & 34.14 & 34.14 & 7.60 & 21.97 & 5258 & 13014 & 541.18 & 541.18 & 1.95 & 40.15\\[0.5mm]
\makebox[0pt][l]{\fboxsep0pt\colorbox{lightgray}{\mystrut\hspace*{1.0\linewidth}}}\!\!
7507 & 10001 & 25.82 & 34.29 & 3.95 & 24.77 & 8025 & 20021 & 157.23 & 157.23 & 0.41 & 41.62\\[0.5mm]
10663 & 14000 & 46.77 & 46.77 & 6.64 & 26.96 & 10011 & 24187 & 179.83 & 280.36 & 1.45 & 60.33\\[0.5mm]
\makebox[0pt][l]{\fboxsep0pt\colorbox{lightgray}{\mystrut\hspace*{1.0\linewidth}}}\!\!
12834 & 17004 & 32.81 & 58.06 & 4.41 & 23.91 & 12409 & 30007 & 114.51 & 114.51 & 0.07 & 55.98\\[0.5mm]
15136 & 20000 & 58.31 & 58.31 & 0.47 & 25.47 & 14691 & 38027 & 731.69 & 1004.13 & 1.40 & 37.81\\[0.5mm]
\makebox[0pt][l]{\fboxsep0pt\colorbox{lightgray}{\mystrut\hspace*{1.0\linewidth}}}\!\!
17554 & 23008 & 45.68 & 64.90 & 0.82 & 25.62 & 17182 & 43013 & 554.27 & 554.27 & 5.59 & 46.36\\[0.5mm]
20590 & 27008 & 51.60 & 76.41 & 4.70 & 26.64 & 20261 & 49021 & 657.47 & 784.02 & 3.52 & 41.41\\[0.5mm]
  
\end{tabular}\end{ruledtabular}

        \caption{In this table we show the initial conditions for the clusters studied in the paper. The ID also corresponds to the initial number of BHs. We also provide the total mass of the cluster, the maximum mass of any single BH in the cluster at t=0 as well as after we have evolved it for the age of the Universe and the radius that encircles half of the total mass of the cluster at both times. The reasons for the maximum mass differing at the beginning and end of the evolution are due to mergers retained in the clusters or the escape of the most massive body.}
    \label{table:ClustersTable}
\end{table*}

For these models to be of any real physical interest Primordial Black Holes (PBHs) should have been created in large numbers at the beginning of the Universe and close enough to each other so that they would have formed clusters so dense as to overcome the Hubble expansion. Several mechanisms could lead to such scenarios~\cite{PBH_Clusters_1, PBH_Clusters_2}. The idea of them making up a significant fraction of the Dark Matter would add relevance to the study but is not essential. To understand the implications of such a scenario, we can estimate the number of clusters needed to get the observed DM mass. Given the clusters' total masses range is $\mathcal{O}(10^3-10^5)\Msun$, we would need around $\mathcal{O}(10^6-10^9)$ clusters to conform the entirety of the dark matter in our galaxy, where we have used a total DM mass of $1.2\times10^{12}\Msun$~\cite{Fragione_2017}. Given the nature of the clusters' components, direct detection is intrinsically difficult; therefore, our approach will focus on identifying potential footprints they may leave behind.

\section{Dynamics}
\label{sec:Dynamics}

The dynamics governing self-gravitating systems are extremely non-linear, thus, trying to infer general qualitative principles is the most we aim to do in this paper. In this section, we will try to shed some light on the understanding of the passage of time for these systems. We will do so by following various global physical variables that would serve as a proxy for the whole cluster. 

\begin{figure}
    \centering
    \includegraphics[width=0.49\textwidth]{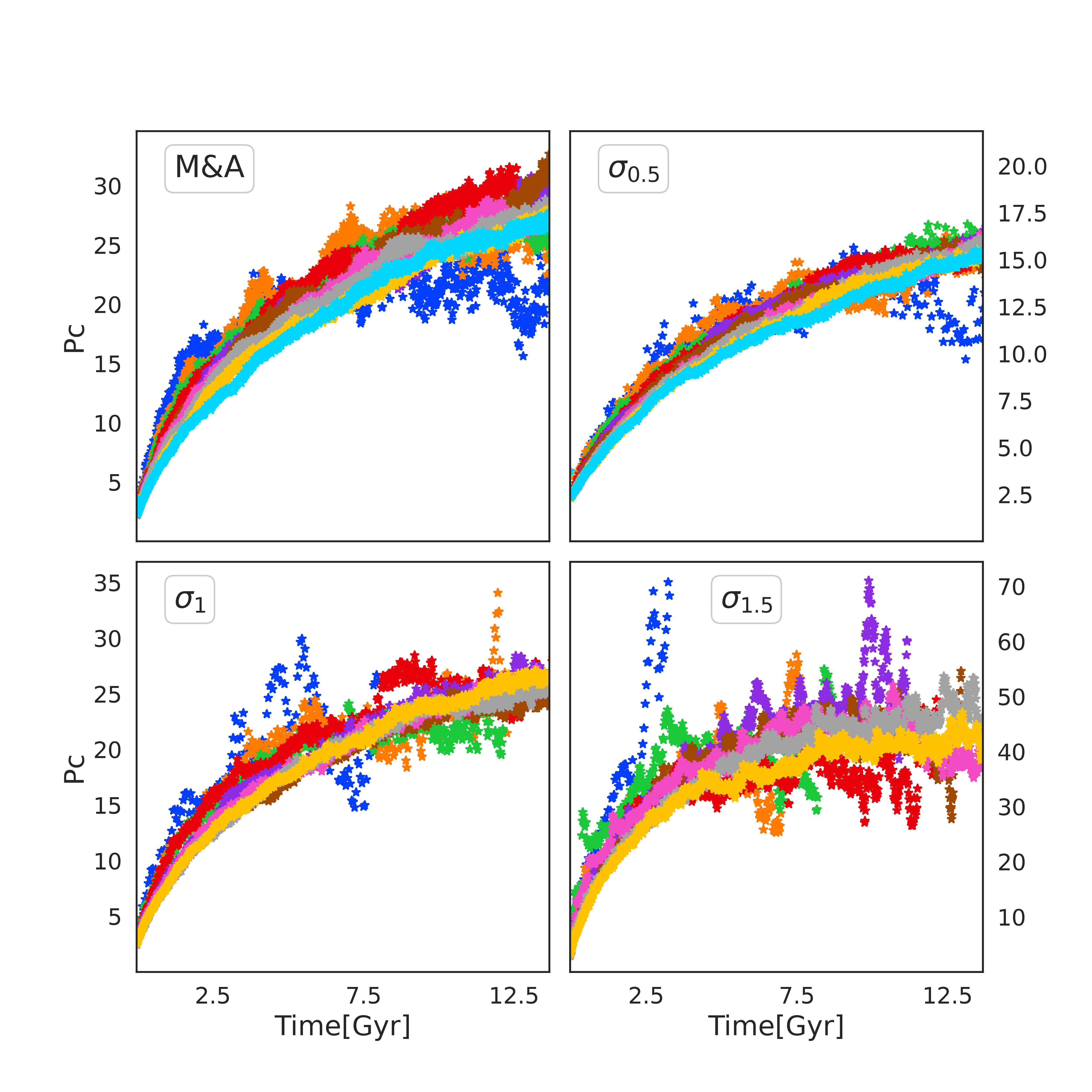}
    \includegraphics[width=0.49\textwidth]{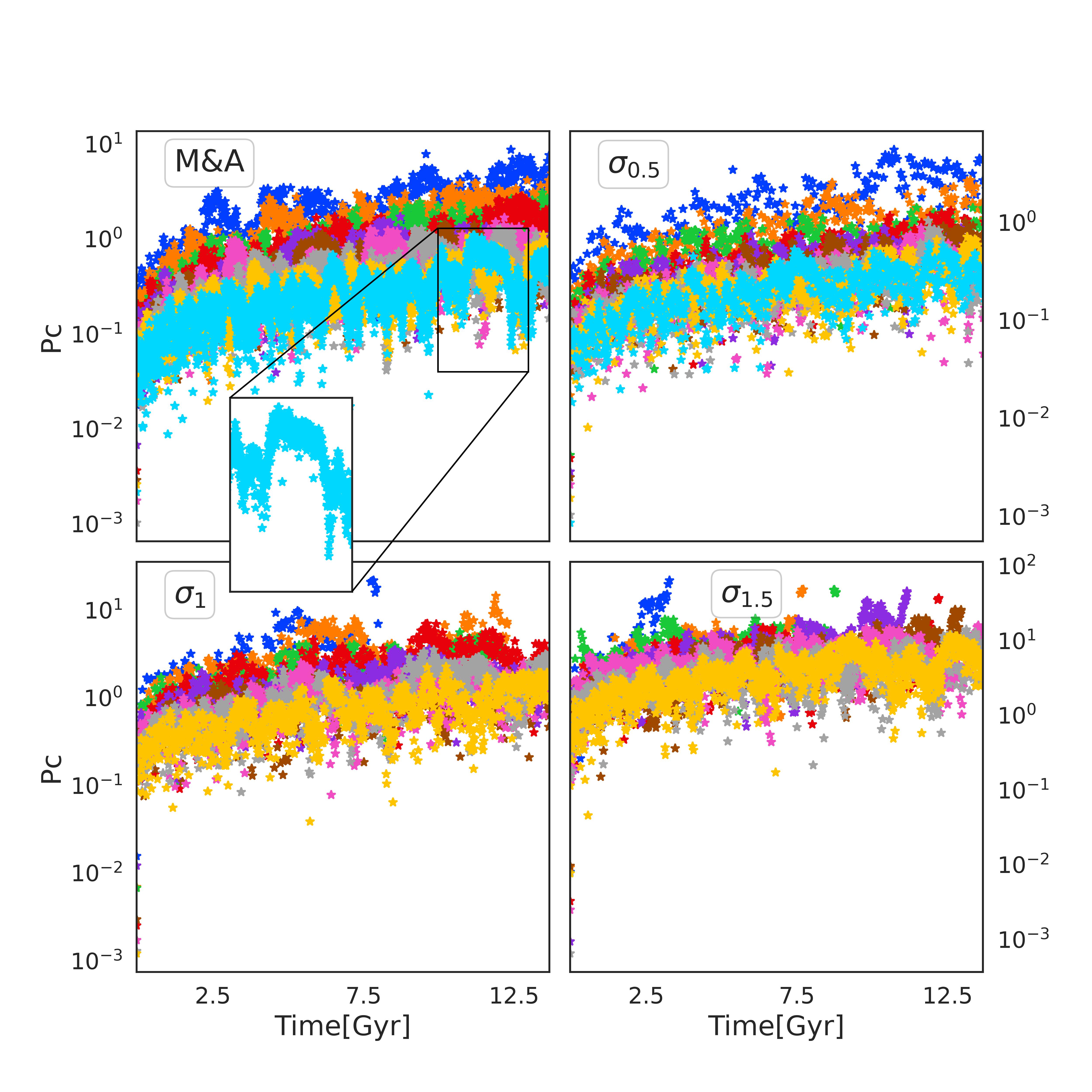}
    \caption{In this figure we plot the Lagrangian radii for 50\% of the mass and the core radii for the various clusters as defined in~\cite{aarseth_2003}. Most lines overlap cluster-wise, so we don't add a legend to distinguish them only pointing out that the lighter clusters are also those showing the largest variance. For the core radii, we include a zooming window to what looks like possible oscillations for the heaviest of the $M\&A$ clusters, although it could also be just shot noise in the calculation.}
    \label{fig:Lagr50}
\end{figure}

From the point of view of the spatial extension of the clusters with time, we can study global quantities such as the Lagrangian radii and the core radii as defined in~\cite{aarseth_2003}. We plot in the left panel of Fig~\ref{fig:Lagr50} the Lagrangian radii for 50\% of the mass of the different clusters. This quantity is also called the half-mass radius. The tendency is for the half-mass radius to grow with time, with a steeper slope at the beginning of the simulation and flattening towards the end, albeit always increasing. In the case of the core radius, which is displayed for all the clusters in the right panel of Fig~\ref{fig:Lagr50}, we observe that it remains constant, with a slight upward trend throughout the evolution. This translates into a faster expansion of the outer layers, converging towards an almost non-expanding core. Additionally, it is interesting to note that most clusters exhibit a similar core radius regardless of the initial number of BHs.

Now we analyse the stability of these clusters. When carrying out the numerical evolution, it is evident from the beginning that BHs are constantly escaping the combined gravitational influence of the cluster and the host galaxy so that the total number being evolved by the code gets reduced with time. In {\tt NBODY6++GPU}, BHs are removed from the cluster evolution once they have reached twice the tidal radius~\cite{TidalRadiusVonHoerner} as calculated using the mass of the cluster at the specific time. This will be our definition of an escaper (single or binary).

\begin{figure}
    \centering\vspace*{-9mm}
    \includegraphics[width=0.46\textwidth]{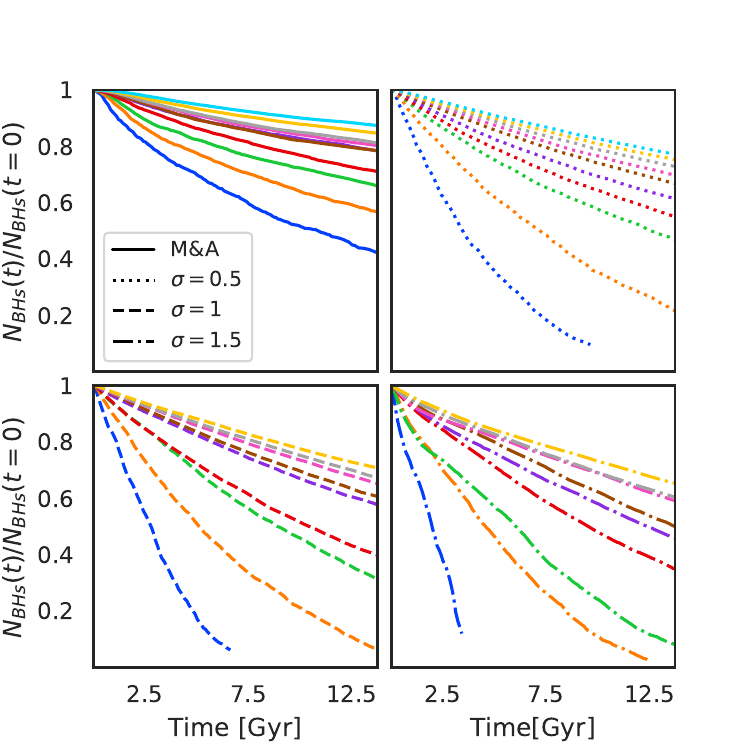}\vspace*{5mm}
    \includegraphics[width=0.52\textwidth]{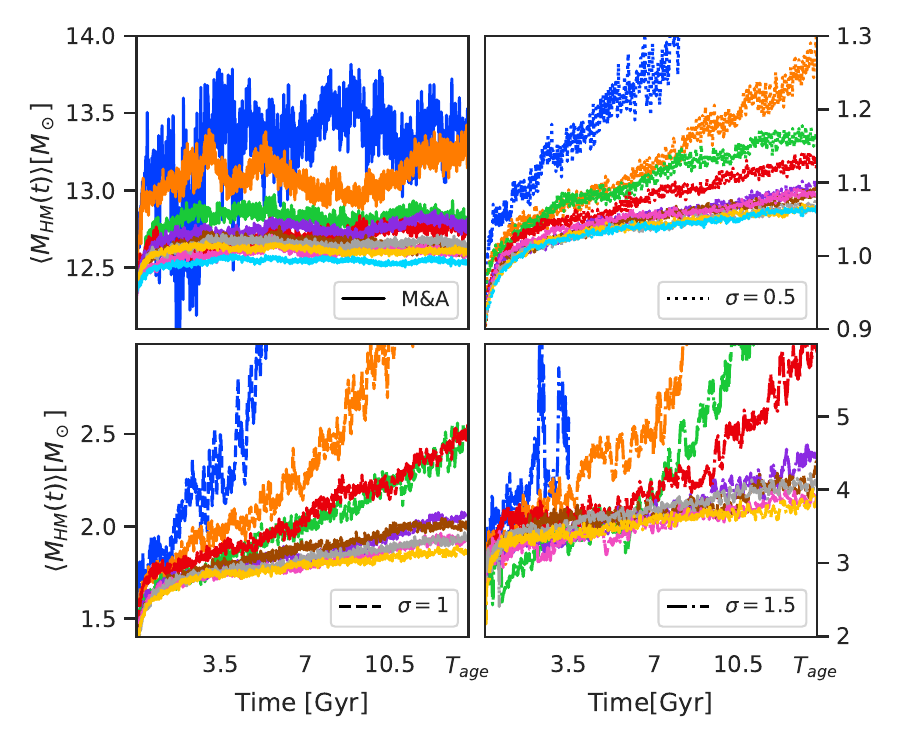}
    \caption{In the left panel of this figure we plot the number of BHs for the various clusters as a function of time normalized to the initial number of BHs. The dark blue lines represent the lightest cluster for a given IMF increasing upwards with no overlapping between the lines in the last $\sim 7$ Gyrs of evolution. In the right panel, we plot the average mass of BHs for the various clusters inside the half-mass radius as a function of time.  }
    \label{fig:NBHswithTime}
\end{figure}

These BHs acquire the needed escape velocity after one or several close encounters with other BHs. This process happens so often that we can consider all clusters to be metastable, that is, most of them don't release enough BHs to dissolve in the age of the Universe, but they will eventually do so. In the left panel of Fig~\ref{fig:NBHswithTime}, we show the number of BHs left in the clusters as a function of time, normalized to the initial number of BHs of each cluster so that the various runs are comparable. We see that some of the lightest clusters of the $\sigma_{i}$ type dissolve completely in the age of the Universe and, in general, the larger the $\sigma$ for the IMF of the cluster, the shorter the expected lifespan. The most stable set of initial conditions seems to be that of $M\&A$. To explain this, we have to understand the main mechanism behind the clusters' dissolution and the difference in their behaviour as a function of the IMF. We can imagine a simplistic scenario in which we just consider 2-body interactions as the main driver of the evaporation of the clusters. This is probably a good approximation as 3+ body encounters are suppressed based on their scattering cross-section. Now, when considering the different IMFs, we see that the main difference across the various $\sigma$ is the mass ratio distribution for random pairs. The larger the $\sigma$ is, the more extreme mass ratios exist, and the expected mass ratio goes further away from 1. In the $M\&A$ case, most of the pairs concentrate close to equal mass ratios. This led us to the conclusion that the more extreme the mass ratios are, the more slingshots of the lighter BHs occur which is translated into more BH escapers. This is also in agreement with the fact that the lighter BHs escape earlier, leaving the heavier ones in the cluster. This can directly be appreciated in the right panel of Fig~\ref{fig:NBHswithTime} where we plot the average mass of the BHs inside the half-mass radius. This quantity is strictly increasing in the $\sigma_i$ cases with a slope that steepens in the last stages of the cluster life due to the wider range of masses present in those clusters. Given the almost monochromatic nature of the $M\&A$ IMFs, the average mass tends to stay constant with a variance that decreases inversely with the initial BH number.

\begin{figure}
    \centering
    \includegraphics[width=0.49\textwidth]{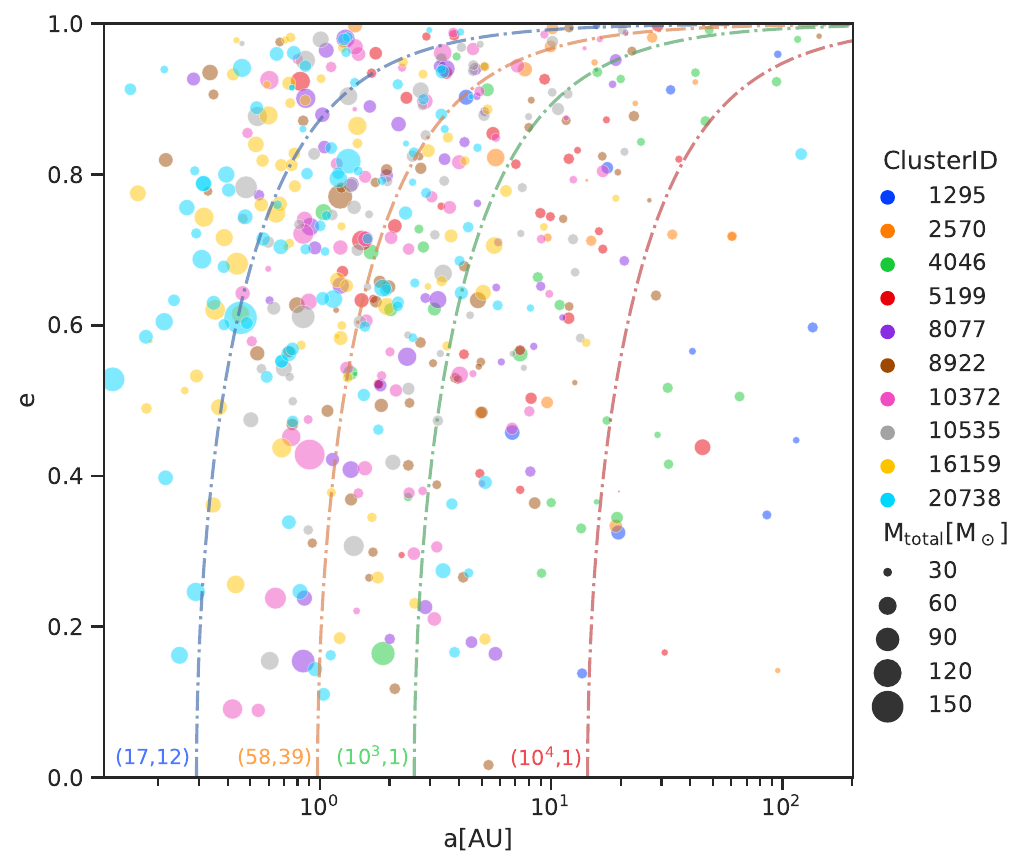}
    \includegraphics[width=0.49\textwidth]{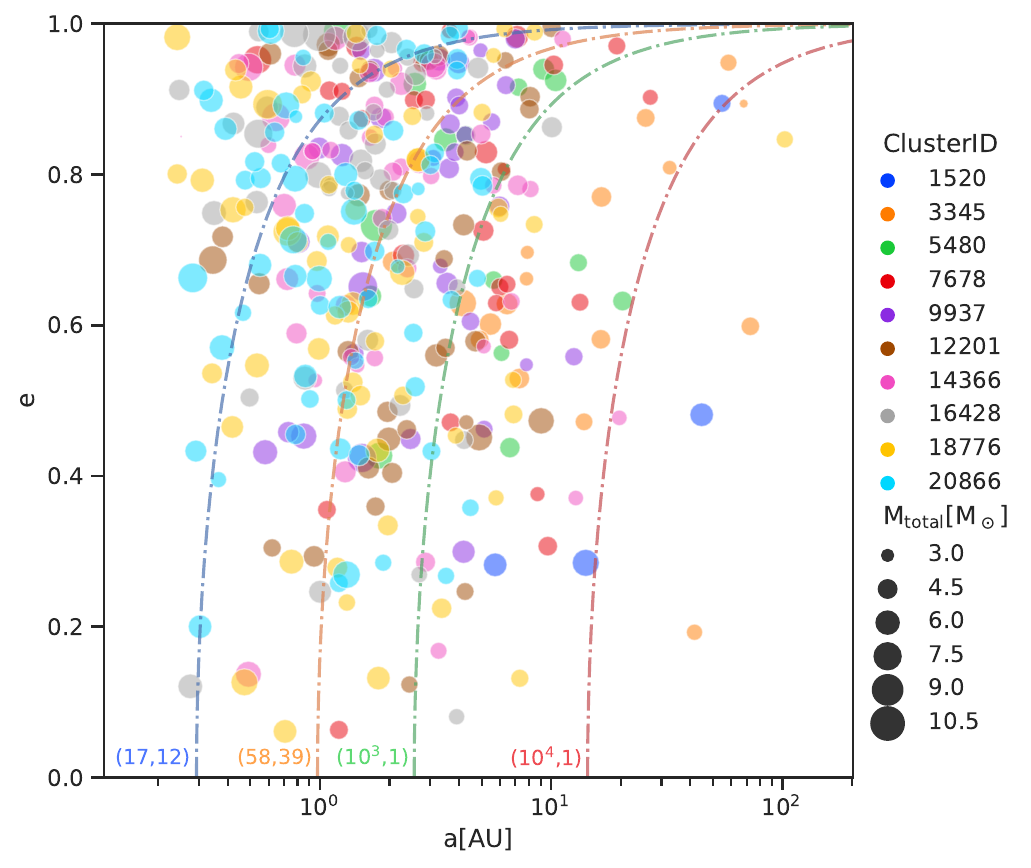}
    \includegraphics[width=0.49\textwidth]{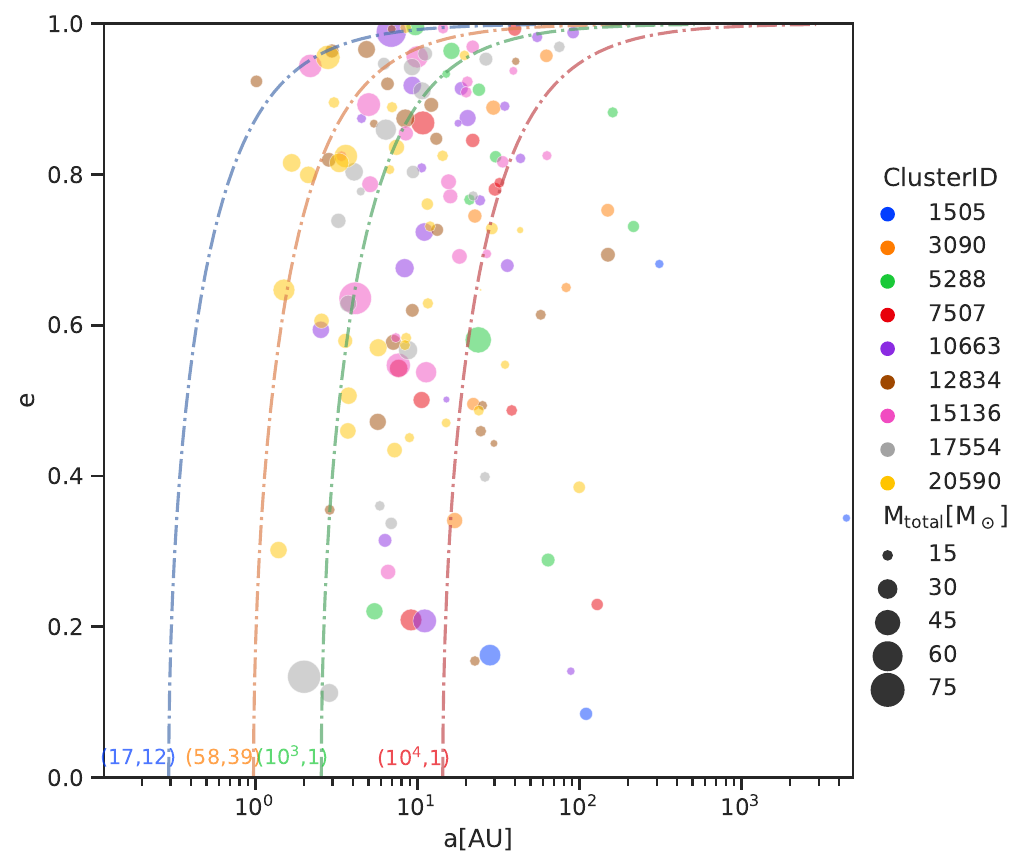}
    \includegraphics[width=0.49\textwidth]{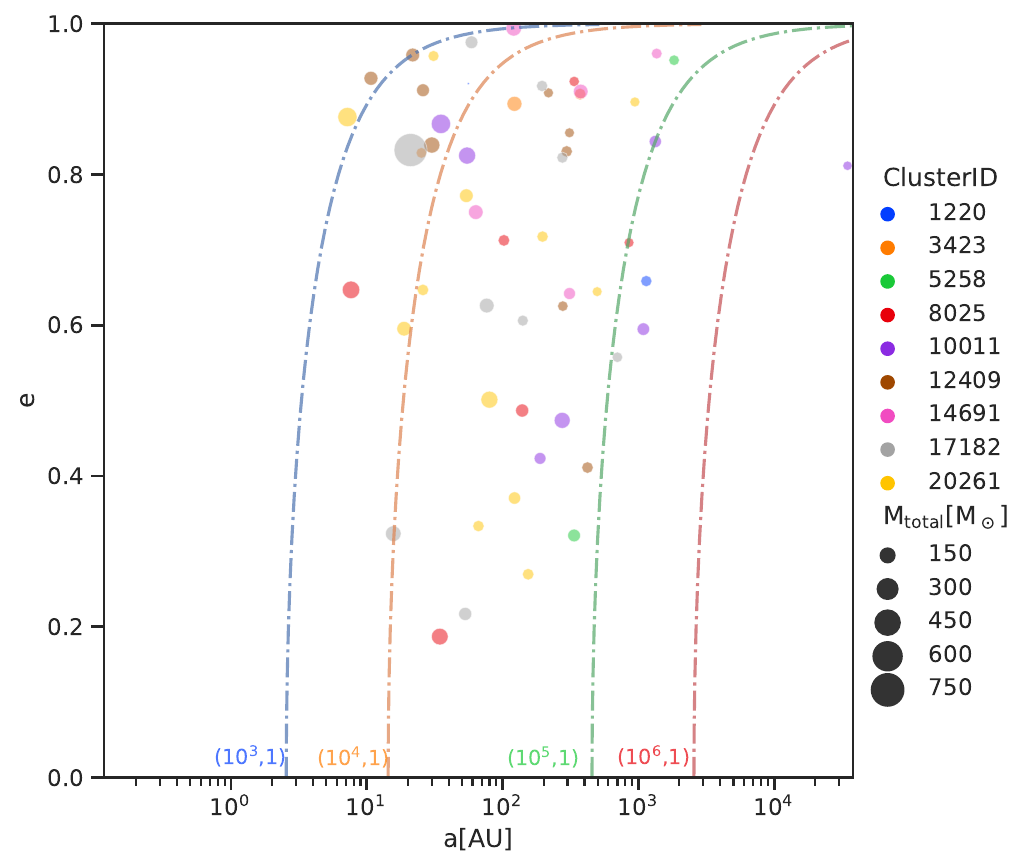}
    \caption{In this figure we plot the initial semimajor axis and eccentricity ($a_0$,$e_0$) of all the BHs that escape the gravitational pull of the clusters within the age of the Universe. These are the binaries logged by the Nbody code as escapers. In order from left to right and top to bottom, we are showing the clusters \{$M\&A$,$\sigma_{0.5}$,$\sigma_{1}$,$\sigma_{1.5}$\}. The dashed lines represent the values of ($a_0$,$e_0$) for which $\tau_{\mathrm{merger}}(m_1,m_2,a_0,e_0) = \mathrm{T}_{Hubble}$ with ($m_1$,$m_2$)[$M_\odot$] being the pairs of number in the base of the plot beside each line. These calculations, however, does not take into account the time delay from the beginning of the simulation until the binary escapes the cluster which can be quite substantial. The width of the dots represents the total mass of the binary defined as $M_{\mathrm{Total}}=m_1+m_2$. }
    \label{fig:SemiMajorEccBinEscapers}
\end{figure}

If a sufficient number of these clusters existed in our Universe, single black hole escapers could potentially be detected from Earth through microlensing or via stellar disruption events. Those that escape as binaries could also be detected in the final stages of the inspiral, as well as the merger and the ringdown in Earth-based gravitational waves interferometers. We will call these binary escapers, off-cluster mergers if the initial conditions regarding their orbits' parameters once they have escaped from the gravitational pull of the cluster are such that they coalesce in the age of the Universe. To calculate the coalescence time, we assume orbit shrinking via gravitational wave emission and use the formula as derived in~\cite{Peters} in post-processing. 

Starting with the single BH escapers, we first acknowledge the fact that the distributions of masses are very similar to the IMF of the clusters with a slight skew towards the lighter BHs due to usual gravitational mass segregation. This means that enough observations of these rogue compact objects would paint a good picture of the progenitor clusters' distributions. 

Regarding binary escapers, of which off-cluster mergers represent a subset, we find a correlation in the distribution of semi-major axes with the initial BH number. The larger the initial number of BHs in the cluster, the tighter the binaries that escape (smaller semi-major axis) and vice-versa. This fact can be visually spotted in figure~\ref{fig:SemiMajorEccBinEscapers}. This is expected as the binaries that may survive in a less dense cluster based on the rate of interactions, may not do so in denser environments. In other words, the binding energy needed for a binary to prevail and escape the cluster grows with the density. Regarding their eccentricities ($e_0$), we see an excess close to maximum eccentricity. This is just the result of the many interactions the binaries need to endure before escaping the cluster.

Results also show that the absolute number of binary escapers increases with the initial BH number within the same type of IMF. Going even further, it also seems to correlate with the stability of the clusters surveyed. The more stable the cluster type, the more binaries escape the gravitational influence with the order from more to less stable being \{$M\&A$,$\sigma_{0.5}$,$\sigma_{1}$,$\sigma_{1.5}$\}.

\section{BBH mergers}\label{sec:Mergers}

During the numerical evolution of the clusters, the centres of two (or more) BHs might get close enough due to the gravitational interactions that their event horizons would merge and the code would start evolving them as a single entity, thus declaring a coalescence. These events occur inside the clusters \cite{DynamicalMergers,Portegies_Zwart_2002}, mainly induced by interactions where 3 or more compact objects are involved~\cite{Heggie_1975}, that is, binary-single interactions or binary-binary interactions. We refer to them as in-cluster mergers. The second type of merger, as described in Section \ref{sec:Dynamics}, occurs when two BHs leave the cluster as a binary bound by their mutual gravitational interaction. Due to gravitational radiation and under the assumption that nothing else ever interacts with these pairs, they will inevitably end up merging. If the time of merger is less than a Hubble time, we refer to them as off-cluster mergers.

\subsection{Merger count}

\begin{figure}
    \centering
    \includegraphics[width=0.54\textwidth]{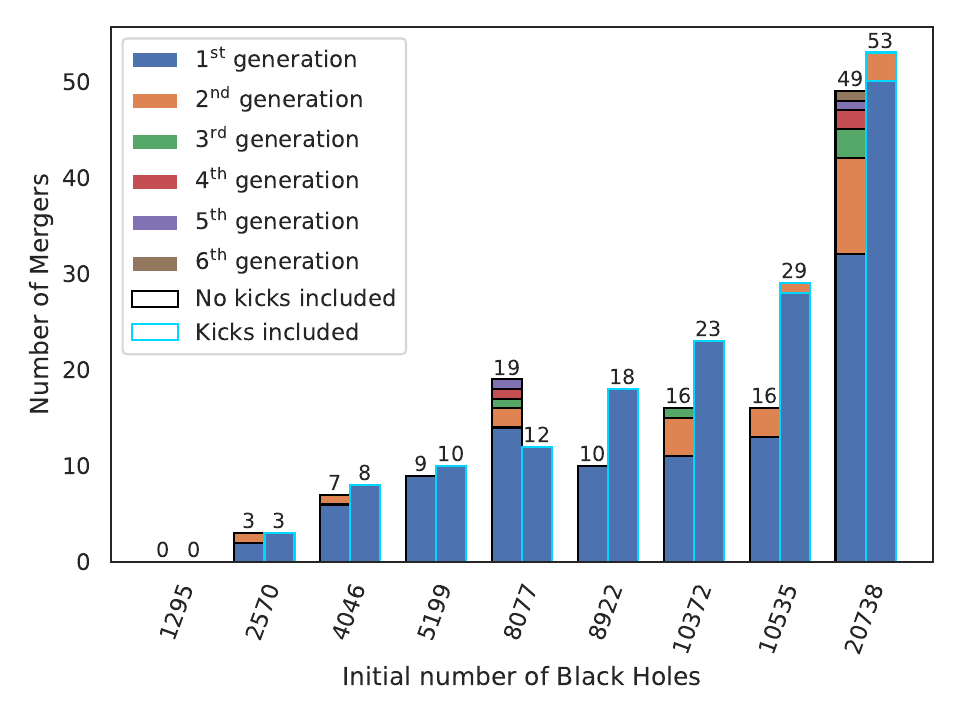}
    \includegraphics[width=0.45\textwidth]{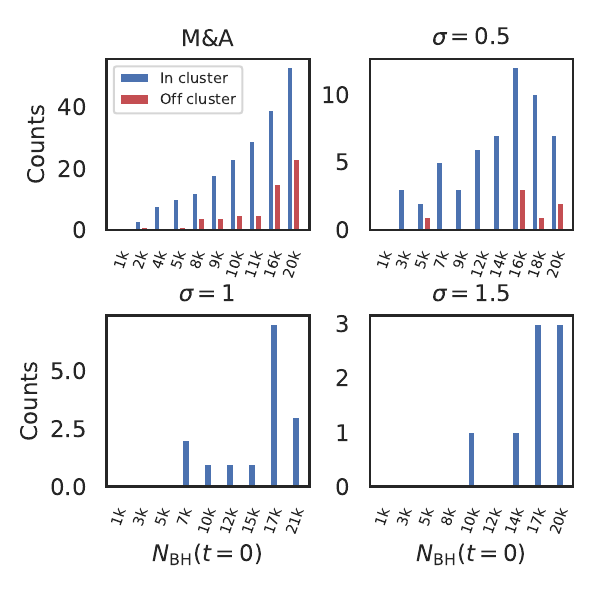}
    \caption{ In the left panel we show a histogram of the number of in-cluster mergers for the $M\&A$ cluster type. The bars with the blue border represent the simulations used throughout the paper, where BH merger kicks are taken into account. The bars with the black border are those same simulations when the kicks are not considered. In the right panel, we plot a histogram of the number of mergers per cluster type. We differentiate, in blue the in-cluster and in red the off-cluster mergers.}
    \label{fig:HierarchichalStudy}
\end{figure}

Understanding the difference between the various initial conditions can be very useful in assessing the underlying mechanisms of the mergers. In the right panel of Fig~\ref{fig:HierarchichalStudy} we show four histograms with the number of mergers per cluster. Discussing first the in-cluster mergers, we find that among the $\sigma_i$ clusters there is a trend in the number of mergers inverse to the width ($\sigma$) of the distribution. In the $\sigma_{0.5}$ and $\sigma_{1}$ cases, a maximum in the merger count seems to arise for $N\sim\mathcal{O}(1.7\cdot10^4)$, although it could be just a statistical anomaly. The total number count for comparable initial $N_{BH}$ lies below that of the $M\&A$ cluster type. This can be explained based on the lower stability of the clusters themselves due to the larger mass range present. The large number of mergers for the $M\&A$ cluster type has as a result also the existence of second-generation BHs, which we discuss in sec~\ref{subsection:HierarchichalMergers}. For this last case, there is a clear upward-sloping tendency in the number count both for in and off-cluster mergers.

We observe a lower number count in the off-cluster merger statistics. This is expected as the delay between the start of the simulations and the ejection of the binary from the clusters can be significant, thus making the expected merger time for the binary escapers way longer than Hubble time. This comes together with the fact that gravitational wave emission is a slow mechanism for shrinking the orbit of the binary in comparison with the interactions governing the in-cluster mergers. Comparing cluster types, we find again fewer counts for the $\sigma_i$ with no off-cluster mergers happening in any $\sigma_{1,1.5}$ clusters. This means that no binary escaper ever merged in the age of the Universe. One of the reasons for this is that the distribution in the semi-major axis of these two cluster types lies at higher values always, which is generally equivalent to larger merger times. Another reason is that the average delay time before escaping is larger for $\sigma_{1,1.5}$ than for the other two clusters. It can also be explained by the fact that there are fewer binary escapers for those two clusters, as can be seen in figure~\ref{fig:SemiMajorEccBinEscapers}.

\subsection{Hierarchical mergers}\label{subsection:HierarchichalMergers}

The product of any merger is another BH with the mass equal to the sum of the masses and spins calculated using~\cite{Jim_nez_Forteza_2017}. These remnants may or may not stay in the cluster as the velocity imprinted on them~\cite{PhysRevD.85.084015,Lousto_2010,PhysRevLett.98.231102} can be very high. Those which not leave the cluster are subject to the possibility of being involved in another merger. To understand how common this phenomenon is, we show a histogram in the left panel of Fig~\ref{fig:HierarchichalStudy} with the total number of BH mergers identified in the lifetime of the clusters for the $M\&A$ cluster type. There are two different bars per Cluster ID. The ones whose border is black represent simulations in which BHs' natal kicks are not taken into account while bars with a blue border are the normal simulations that do include such a physical phenomenon and thus represent a more complete description of reality. The initial conditions are the same for both cases. We find that when kicks are not present, the probability that a remnant stays in the cluster is high as the velocity after the merger is obtained based on linear momentum conservation. However, when kicks are included, relativistic effects imprint a velocity in the remnant that is usually much larger than the escape velocity of the cluster. This has as a consequence a sharp drop in the probability of hierarchical mergers. We can see this reflected in the fact that hierarchical mergers of up to $6^{\mathrm{th}}$ generation can happen in most of the clusters if kicks are ignored but it's only for N larger than 10535 BHs that we find hierarchical mergers when kicks are considered. What is more, anything beyond $2^{\mathrm{nd}}$ generation is not on the cards, as the kick velocity is even larger in the case where progenitor spin is non-zero~\cite{Campanelli_2007}. This has broader consequences in the distribution of other binary parameters, such as the final spin. As can be seen in Fig.~\ref{fig:SpinDistribution} a limited number of BH remnants have $a_f>0.7$ and they are all produced via second-generation mergers. A variation in the statistics of such products would lead to critical differences in the detected population of BHs via for example gravitational wave interferometers. 

This result is of great importance as it highlights the significant impact that an occasionally overlooked physical effect can have on observational probes. Any simulation lacking the modelling of BH merger kicks is doomed to overestimate the hierarchical merger rate and the skew to larger remnant spins. 

\subsection{Merger Rate}
\label{subsection:MergerRate}

\begin{table}[htbp]
\centering
\begin{tabular}{ccc|ccc}
\toprule
\hline\\[-3mm]
\textbf{Cluster ID} & \textbf{Redshift} & \textbf{Rate z=0} & \textbf{Cluster ID} & \textbf{Redshift} & \textbf{Rate z=0} \\
 \textbf{$M\&A$ } & \textbf{last merger} & \textbf{(events/yr/Gpc$^3$)} & \textbf{$\sigma_{0.5}$ } & \textbf{last merger} & \textbf{ (events/yr/Gpc$^3$)} \\
\hline\\[-3mm]
\midrule
2570 & 5.89 & 0.00 & 3345 & 1.71 & 0.00 \\
\makebox[0pt][l]{\fboxsep0pt\colorbox{lightgray}{\mystrut\hspace*{0.7\linewidth}}}\!\!
4046 & 2.79 & 0.00 & 5480 & 0.03 & 263.69 \\
5199 & 0.18 & 1.36 & 7678 & 2.67 & 0.00 \\
\makebox[0pt][l]{\fboxsep0pt\colorbox{lightgray}{\mystrut\hspace*{0.7\linewidth}}}\!\!
8077 & 0.44 & 2.72 & 9937 & 0.83 & 80.93 \\
8922 & 1.77 & 0.00 & 12201 & 1.65 & 0.00 \\
\makebox[0pt][l]{\fboxsep0pt\colorbox{lightgray}{\mystrut\hspace*{0.705\linewidth}}}\!\!
10372 & 0.43 & 1.36 & 14366 & 0.33 & 77.48 \\
10535 & 0.04 & 5.45 & 16428 & 0.60 & 29.55 \\
\makebox[0pt][l]{\fboxsep0pt\colorbox{lightgray}{\mystrut\hspace*{0.705\linewidth}}}\!\!
16159 & 0.18 & 6.81 & 18776 & 0.34 & 45.33 \\
20738 & 0.05 & 14.98 & 20866 & 0.44 & 18.34 \\
\hline\\[-3mm]
\bottomrule
\end{tabular}\caption{ We present the local merger rate and redshift of the last identified in-cluster BH merger for every cluster with at least one merger within the age of the Universe. We focus on the \{$M\&A$,$\sigma_{0.5}$\} cluster types due to their larger merger counts. The rate is calculated using all the in-cluster mergers with $z<1$ to approximate the local merger rate, assuming that all of the Dark Matter is contained within such dense clusters.}
\label{tab:combined_clusters}
\end{table}

\begin{figure}
    \centering
    \includegraphics[width=0.5\textwidth]{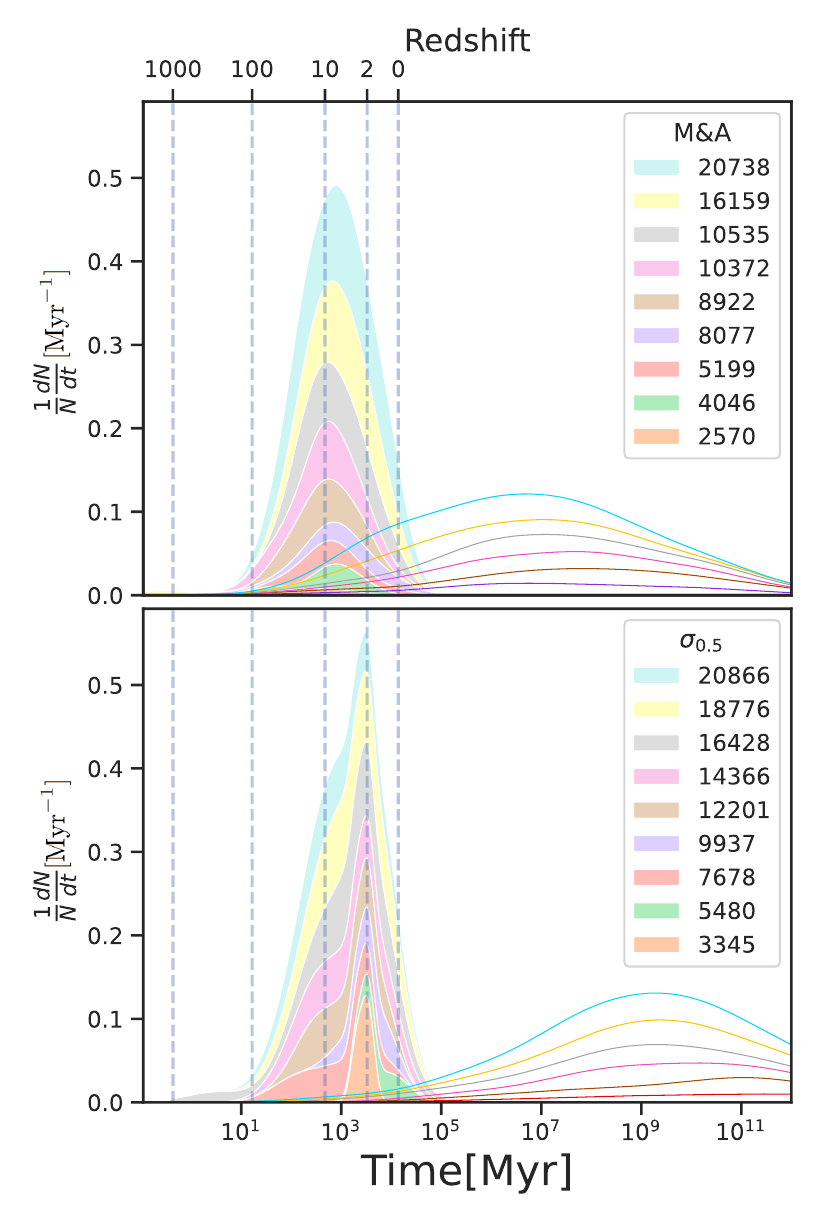}
    \caption{We show the in-cluster and binary escapers merger rate for the cases of $M\&A$ and $\sigma_{0.5}$, which have the highest number of mergers, allowing for more accurate kernel density plots of the underlying merger rate. The binary escapers' merger rates are represented as solid lines with similar colours as the filled density plots depicting the in-cluster merger rates. We observe a general trend where the in-cluster mergers peak around 0.54 Gyr after the Big Bang for $M\&A$ clusters and 3.3 Gyr for $\sigma_{0.5}$ clusters, well before a Hubble time, while the binary escapers' merger rate extends significantly beyond it.
    }
    \label{fig:MergerswithTime}
\end{figure}

Following the merger history of the various clusters throughout their evolution could give us clues about possible probes of these models, as currently, our best chance to infer the existence of these objects is via gravitational waves. For that purpose, we plot in Fig~\ref{fig:MergerswithTime} the in-cluster and off-cluster merger rates as a function of time for the cluster types \{$M\&A$,\,$\sigma_{0.5}$\} based on their larger merger statistics compared to the other two. The first thing we observe is there is a pronounced peak for the in-cluster merger rate at a time $T\sim\{0.5,\,3\}~\mathrm{Gyrs}$ or $z\sim\{9,\,2\}$ respectively for the two cluster types and a distribution for the off-cluster merger rate very broad with time. In Table~\ref{tab:combined_clusters} we include the estimated local merger rate of the different clusters assuming they individually comprise all of the dark matter in our Universe. This way of presenting the rates for individual clusters rather than cumulatively for all clusters within the same type facilitates comparison with the rate we could estimate if we observed any one of these clusters in our Universe. Given the limited statistics available in some cases, these estimates are susceptible to shot noise for the single clusters.

From the in-cluster mergers perspective, current Earth-based interferometers are on the verge of being able to detect events coming from as far as $z=2$, with examples in O3b going as high as $z=1.18^{+0.73}_{-0.53}$ in the case of GW190403$\_$051519~\cite{GWTC3}. Current star formation models~\cite{annurev-astro-081811-125615} predict a peak in the merger rate corresponding to the peak in the star formation rate ($z\sim1.8$ = Madau-Dickinson 3.5 Gyr after BB) plus a time delay of about a Gyr due to the time of BH collapse, binary formation and subsequent merger ($z\sim1.1-0.9$). Finding an excess beyond this redshift would possibly hint towards one of the models of dense clusters of PBH analysed here. 

In the case of off-cluster mergers, the rate is smaller, and instead of increasing, their redshift dependence decreases towards larger redshifts. This characteristic could be utilized to distinguish between their primordial and astrophysical origins.

\subsection{Mass distribution}

\begin{figure}
\centering
    \includegraphics[width=0.5\textwidth]{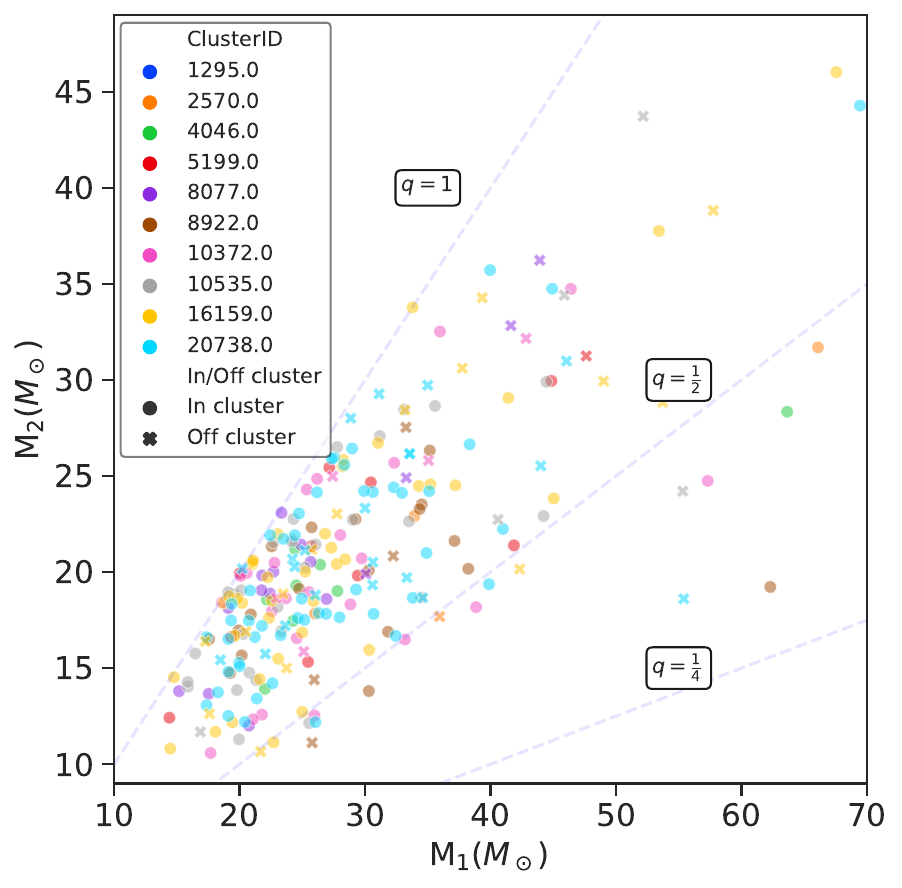}
    \includegraphics[width=0.49\textwidth]{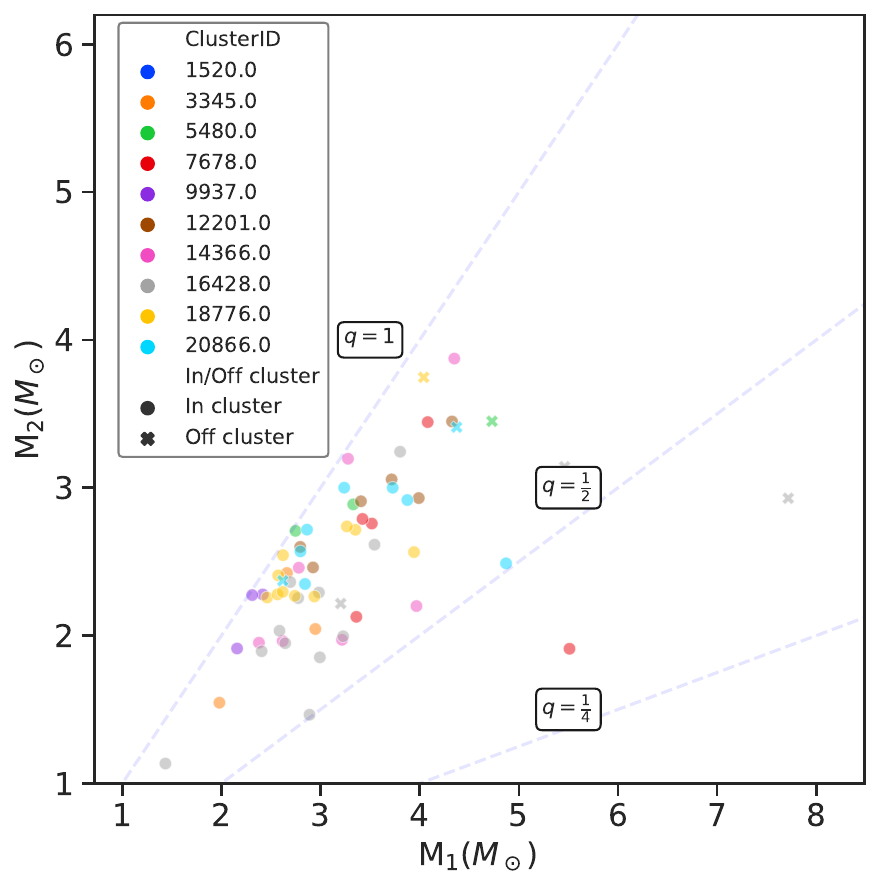}
\caption{In this figure, we present the distribution of the masses of the BHs that merge within the age of the Universe. They are categorized into two groups: those merging inside the clusters, identified as collisions by the code (in-cluster), and those accounted for as binary escapers and evolved outside of the simulation (off-cluster). Some clusters don't produce any merger and thus, not every color in the legend is represented in the plots. The left plot corresponds to the $M\&A$ type and the right plot represents the $\sigma_{0.5}$ models.}
\label{fig:M1M2MergerDist}
\end{figure}

Another approach to analyzing BBHs is through their mass distribution, which is highly dependent on the original distribution of masses due to the strong suppression of hierarchical mergers in realistic simulations. In Fig~\ref{fig:M1M2MergerDist} we present a scatter plot of the mergers in the ($M_1$,$M_2$)[$M_\odot$] plane. We can differentiate the two cases: in-cluster mergers directly identified by the code and depicted as circles in the figure and off-cluster mergers that we evolve in post-processing. A noticeable clustering around equal mass ratios is observed, deviating from the expected binary distribution one would obtain by randomly drawing pairs from the initial mass function. This suggests a genuine preference for $q=1$ embedded in the mechanisms responsible for these mergers. This preference implies that the binary formation cross-section has a maximum for equal mass ratios, while unequal ratios result in more hyperbolic encounters. The BBH merger with the minimum mass ratio identified sits around $\sim 0.2$ in the $\sigma_{1.5}$, not shown in the plot, which is above the minimum mass ratio detected in a confident BBH which is $\sim 0.1$ for GW190814.

\subsection{Spin distribution}

Initially, the spins of all the BHs are set to 0 inspired by the isotropic collapse of the primordial overdensities predicted by most PBH formation theories. This assumption, nonetheless, may also apply to stellar BHs as the efficiency of the angular momentum transport from the spin of the progenitor star is still under debate (e.g. \cite{AngularMomentumTransportI,AngularMomentumTransportII,AngularMomentumTransportIII}). The various BHs can acquire spin via merging with other BHs. The final spin of the remnant is calculated using \cite{Jim_nez_Forteza_2017}, which depends also on the progenitors' spins and masses. It is also possible that due to the multiple close encounters in dense clusters, the spin of the BH population is induced stochastically, with a final spin distribution peaked at zero with a dispersion of about $\sigma_s=0.2$~\cite{Jaraba:2021ces}. We did not consider in this work the induced spin due to close hyperbolic encounters in the distribution of spins of the BH population.

\begin{figure}
    \centering
    \includegraphics[width=0.99\textwidth]{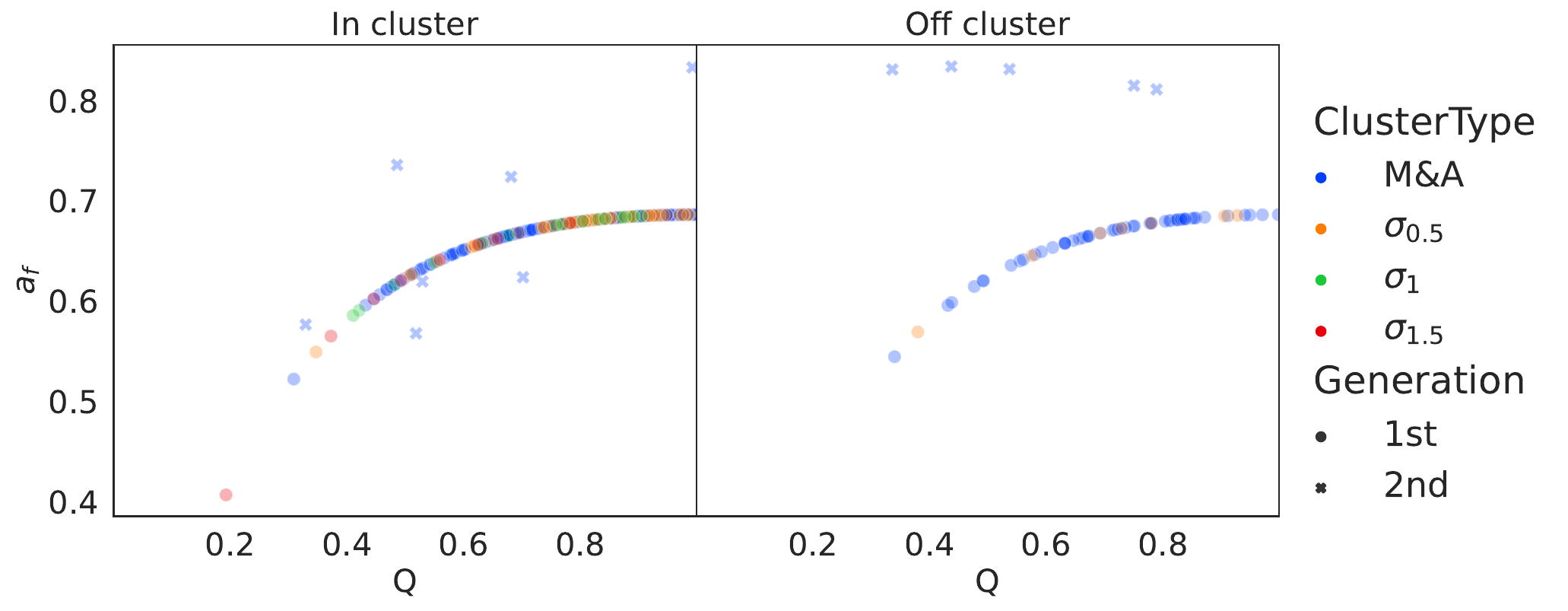}
    \caption{In this plot we show the value for the final spins of the remnant BHs as a function of the binary mass ratios of the progenitors. The simulations assume zero spins as the initial condition. We distinguish between mergers happening inside the cluster and outside and also between 1st and 2nd generation mergers. Spins are calculated using \cite{Jim_nez_Forteza_2017} either by the code itself or in post-processing.}
    \label{fig:SpinDistribution}
\end{figure}

In Fig~\ref{fig:SpinDistribution} we present the distribution for the absolute magnitude of the spin of the remnant BHs. The left panel illustrates the distribution for in-cluster mergers, while the right panel depicts the same for off-cluster mergers. We observe that the bulk of the distribution is similar for both cases, as the initial spins for all BHs are 0, and thus, the final spin depends entirely on the initial masses. The most frequent value for the final spin is around $a_{\mathrm{f}}=0.68$ as you need to reach very unequal mass ratios $q<0.4$ to go below $a_{\mathrm{f}}=0.6$. There are however, some extreme spins, sitting around $a_{\mathrm{f}}=0.9$ due to $2^{\mathrm{nd}}$ generation mergers where one of the progenitors already had a non-negligible spin. They represent a $4\%$ of all the BBHs that coalesce while representing a larger fraction of the total number of off-cluster mergers. The smallest final spin sits around $a_{\mathrm{f}}=0.4$, corresponding to an extreme mass ratio of $q=0.2$ in the $\sigma_{1.5}$ type of clusters and the largest one lays very close to $a_{\mathrm{f}}=0.9$, coming from a second generation merger and high total mass. This distribution aligns with general observational results and our expectations that most, if not all, BHs in our universe are Kerr BHs, as there are many mechanisms to gain spin but very few to lose it.

\subsection{Residual eccentricity distribution}

Another parameter crucial for understanding the nature of BBHs is the residual eccentricity of the binaries. Looking at fig~\ref{fig:MergerswithTime}, we can differentiate the in-cluster and the off-cluster mergers as two very different cases. The in-cluster mergers are the result of complex dynamical processes where more than two bodies are usually involved. The main process by which the BHs end up merging is not gravitational wave emission, and thus, we expect to find highly eccentric signals as eccentricity cannot be radiated away. This could be a distinct feature if we were to detect an eccentric merger at a sufficiently high redshift as we would not expect in-cluster mergers to happen before the peak star formation rate era. This kind of highly eccentric merger, however, presents challenges on its own to be measured (see~\cite{PhysRevLett.126.201101}).  Given the lack of a significant inspiral phase in these events, in-cluster mergers would likely manifest in Earth-based gravitational wave interferometers as short bursts, of which one example might be GW190521~\cite{GW190521}. This comes with various challenges as the very low number of detectable cycles leaves the floor open to numerous alternative hypotheses~\cite{PhysRevLett.126.081101,gayathri2022eccentricity,Romero_Shaw_2020}.

\begin{figure}
    \centering
    \includegraphics[width=0.78\textwidth]{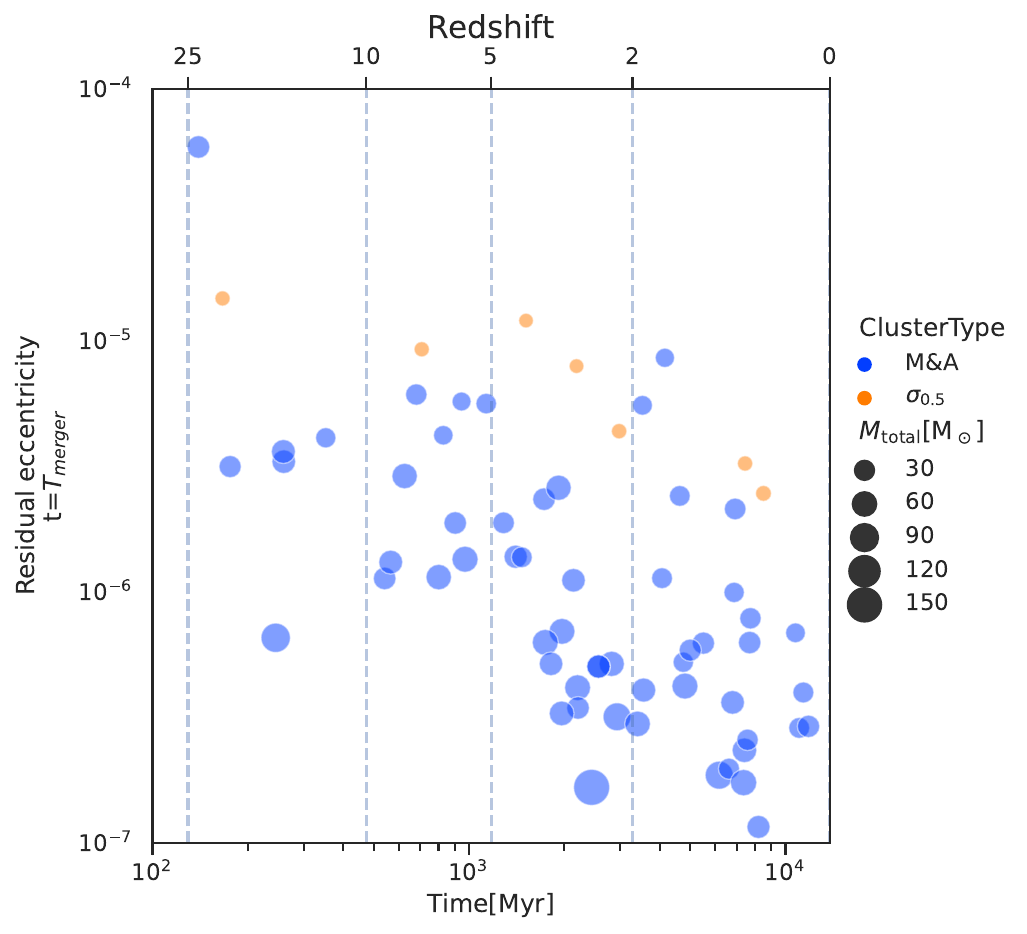}
    \caption{In this figure, we display the residual eccentricity of the off-cluster mergers. We define residual eccentricity as the eccentricity the binary would possess at the time of merger, after it has been radiated away throughout the inspiral phase. This residual eccentricity could potentially be detected from ground-based interferometers. }
    \label{fig:ResidualEccentricity}
\end{figure}

The other case, the off-cluster mergers, follow a completely different route and the only mechanism by which they end up merging is via orbit shrinking through GW emission. As a result, eccentricity is radiated away~\cite{Peters} before the binary reaches the detectability threshold in current Earth-based interferometers. To investigate the validity of such a hypothesis, we plot in fig~\ref{fig:ResidualEccentricity} the residual eccentricity for the off-cluster mergers, that is, the eccentricity that, based on quadrupolar GW emission, the binaries would have at the moment of the merger. Most of the cases that would concern us ($z<2$) present a maximum residual eccentricity of $e\sim10^{-4}$, which is hardly detectable given current sensitivity and the fact that it might be confused with spin effects~\cite{Romero_Shaw_2023}. The detectability threshold lies at around $e\sim10^{-3}$ for a high enough SNR event, justifying the use of the quasi-circular approximation.

\section{Conclusions}
\label{sec:Conclusions}

In this paper, we have reported the results of our study of the phenomenology of BHs-only clusters with different initial mass functions based on both astrophysical and cosmological assumptions for their origin. 

From a dynamics point of view, we have encountered differences in the evaporation as well as the expansion rate of the clusters. The main distinction between cluster types can be understood via their stability, which seems to depend highly on the mass ratios of the typical encounters occurring inside of them and driving the energy exchange between the BHs. In this regard, we find that single BH escapers' distributions resemble very closely the original IMF except for a larger skew towards smaller masses and the total number of binary escapers seems to be correlated again with the stability criteria so that, the more stable the cluster is, the more binaries can fly from its gravitational potential well.

Regarding BBH coalescences, we distinguished between in-cluster and off-cluster mergers due to their significant differences in orbital parameters and potential detectability from Earth. On the one hand, in-cluster mergers are produced via 3+ body encounters inside the densest regions of the clusters, with a peak in the merger rate at a relatively high redshift, generally above the detectability horizon of current interferometers, although this is highly model-dependent. These mergers are also characterized by very high residual eccentricity and few detectable cycles from Earth. On the other hand, off-cluster mergers occur outside of the cluster due to gravitational wave emission. This mechanism leads to a very low residual eccentricity ($e<10^{-4}$), leading to quasi-circular BBH coalescence as observed from Earth. The masses of the BBH mergers depend very much on the IMF although we find a trend favouring mass ratios close to 1 independent of the IMF. Given our zero-natal spin assumption, most of the final spins depend on the progenitor masses with a very low probability of non-zero progenitor spin ($<4\%$).

We also conducted a comparison between the $M\&A$ clusters with and without considering BH merger kicks, finding a strong dependence on the existence and number of hierarchical mergers. This finding provides valuable insight into the accuracy of numerical simulations. Predictions regarding the occurrence of these types of mergers cannot be relied upon if merger kicks are not incorporated.

In summary, this work aims to elucidate the phenomenology of complex non-linear systems represented by gravitationally-bound BH-only clusters. We hint at some results as promising candidates for unexplained observations and expand the world of possibilities in numerical research.

\section*{Acknowledgements}

We would like to thank Rainer Spurzem and Manuel Arcca Sedda for their valuable comments and suggestions during the drafting of this manuscript.
The authors acknowledge the use of the publicly available codes: {\tt NBODY6++GPU}. 
They acknowledge support from the research project  PID2021-123012NB-C43 and the Spanish Research Agency (Agencia Estatal de Investigaci\'on) through the Grant IFT Centro de Excelencia Severo Ochoa No CEX2020-001007-S, funded by MCIN/AEI/10.13039/501100011033. JFNS acknowledges support from MCIN through Grant No. PRE2020-092571. The authors acknowledge the use of the Hydra cluster at the IFT, on which some of the numerical computations for this paper took place.


\bibliography{Refs}

\end{document}